# Deep Learning Based Automated COVID-19 Classification from Computed Tomography Images

Kenan Morani*, Devrim Unay**

*kenan.morani@gmail.com, **devrim.unay@idu.edu.tr

Electrical and Electronics Engineering Department, Izmir Democracy University, Izmir, Turkey

*Abstract—* *A method of a Convolutional Neural Networks (CNN) for image classification with image preprocessing and hyperparameters tuning was proposed. The method aims at increasing the predictive performance for COVID-19 diagnosis while more complex model architecture. Firstly, the CNN model includes four similar convolutional layers followed by a flattening and two dense layers. This work proposes a less complex solution based on simply classifying 2D-slices of Computed Tomography scans. Despite the simplicity in architecture, the proposed CNN model showed improved quantitative results exceeding state-of-the-art when predicting slice cases. The results were achieved on the annotated CT slices of the COV-19-CT-DB dataset. Secondly, the original dataset was processed via anatomy-relevant masking of slice, removing none-representative slices from the CT volume, and hyperparameters tuning. For slice processing, a fixed-sized rectangular area was used for cropping an anatomy-relevant region-of-interest in the images, and a threshold based on the number of white pixels in binarized slices was employed to remove none-representative slices from the 3D-CT scans. The CNN model with a learning rate schedule and an exponential decay and slice flipping techniques was deployed on the processed slices. The proposed method was used to make predictions on the 2D slices and for final diagnosis at patient level, majority voting was applied on the slices of each CT scan to take the diagnosis. The macro F1 score of the proposed method well-exceeded the baseline approach and other alternatives on the validation set as well as on a test partition of previously unseen images from COV-19CT-DB dataset.*

## 1. INTRODUCTION

The COVID-19 virus, or the severe acute respiratory syndrome coronavirus 2 (SARS-CoV-2), is believed to have initially originated from the species of bats and transmitted to human beings in December 2019. The virus spread rapidly all around the world, affecting lots of people and claiming lives [1]. COVID-19-infected individuals have experienced fever on the onset, generalized fatigue, dry coughing, diarrhea among other possible symptoms [2]. Early detection and isolation are vitally important to successfully handle the COVID-19 pandemic. Studies have shown the importance of lung imaging for that cause [3]. With that, automated solutions were proposed for COVID-19 detection through medical images such as computed tomography (CT) scans using different algorithmic methods [4].

The proposed methods would report classification performance scores in different matrices including accuracy, precision, recall, specificity, and F1 scores [5]. In case of an unequal number of observations in the classes (unbalanced data), accuracy of the solutions is important, but it might be misleading. If this is the case, then the model can be assessed in terms of its "Precision" and "Recall". If the former is high, then that means the model gives more relevant results than irrelevant ones. On the other hand, if the latter is high then that means the model gives most of the relevant results (whether irrelevant ones are also returned). Therefore, for unbalanced classification problems, the weighted average of the two scores or the macro F1 score can be used to evaluate the classification performance of a model in a more reliable manner [6].

In this paper, the macro F1 score was used to compare the performances of different deep learning models and methods validated on the same dataset. The comparison was made at two levels: slice-level and patient-level. Our deep learning model resulted in state-of-the-art macro F1 score at slice level. This led to the first comparison with other models and methods at both slice level or patient level, according to the highest macro F1 score reported. Results at patient level was obtained by combining the deep learning model with two main processing. The processes can be referred to as slice processing and hyperparameter tuning. The final method achieved a macro F1 score at patient level exceeding a baseline score and many other alternatives on the COV19-CT-DB database.

The design of those models/methods is aimed at finding an automated solution for COVID-19 diagnosis via CT-scan images. The proposed classification solution in this paper progresses from a deep learning model consisting of four similar 2D convolutional layers followed by a flattening layer and two dense layers to a method that is then used to make diagnosis predictions at patient's level using different thresholds via class probabilities and voting from the slices.

The main contributions of this work can be listed as follows:

- We present a less complex deep neural network to achieve COVID-19 diagnosis from CT images.

- We show that processing CT images with a Region of Interest (ROI) dedicated to the lung region improves diagnostic performance.
- We propose taking patient-level diagnosis from slice-level processing via the proposed method.
- We evaluate the performance of the proposed solution on a recent, relatively large, challenging dataset.

## 2. RELATED WORK

Recently, deep Transfer learning and Customized deep learning-based decision support systems are proposed for COVID-19 diagnosis using either CT or X-ray modalities [7, 8, 9]. Some of these systems are developed based on pre-trained models with transfer learning [10, 11], while a few others are introduced using customized networks trained from scratch [12, 13, 14].

One approach proposed a novel COVID-19 lung CT infection segmentation network, named Inf-Net [13]. The work utilized implicit reverse attention and explicit edge-attention aiming at identification of infected regions in CT images. The work also introduced a semi-supervised solution, Semi-Inf-Net, aiming at alleviating shortage of high-quality labeled data. The proposed method was designed to be effective in case of low contrast regions between infections and normal tissues.

Another approach used deep learning based automated method validated using chest X-ray images collected from different sources. Different pre-trained CNN models were compared, and the impact of several hyperparameters was analyzed in this work. Finally, the best performing model was obtained. ResNet-34 model outperformed other competitive networks and thus development of effective deep CNN models (using residual connections) proved to give a more accurate diagnosis of COVID-19 infection [14].

A recent work introduced an ensembled deep neural network (IST-CovNet), providing evaluation of different 2D and 3D approaches on two different datasets and discussing the effects of preprocessing, segmentation, and classifier combination steps on the performance of the approach [15]. The final model combined the use of a novel attention mechanism with slice level combination using LSTMs (Long Short-Term Memory) and an extended architecture for 3D data. This approach proven to increase accuracies in both 2D and 3D models validated on the public dataset "MosMedData" [16], achieving state-of-the-art performance. Furthermore, the authors introduced a large, collective dataset referred to as "IST-C", which was made public to contribute to the literature. Their approach also proven to give high performance on their introduced dataset.

While there are multitude of studies aiming at COVID-19 diagnosis using different dataset/dataset combinations, we focused a recent, heterogeneous database of CT scan images, called "COV19-CT-DB". Our work was employed on this particular database for different reasons. 1) COV19-CT-DB is a recent database shared via an international competition about mid of the year 2021 and was used by several international teams for COVID-19 diagnosis. 2)The CT images in the database are challenging. In simpler terms, both the number of slices and the observed anatomy in the 3D field-of-view in each CT-scan varies - as elaborated in the dataset section. This variability further increases the challenging nature of the problem.

Using the COV19-CT-DB series of images [17], a baseline approach introduced a deep neural network, based on CNN-Recurrent Neural Network (RNN) architecture. The CNN part of the model extracts features from the images while the following RNN part takes the final diagnostic decision [18, 19, 20].

Another study, which used the same database (COV19-CT-DB) for validation, introduced a different method [21]. In this study 2D deep CNN models were trained on individual slices of the database. Performances of the following pre-trained models were compared -VGG, ResNet, MobileNet, and DenseNet. Evaluation of the models was reported both at slice level (2D) as well as at patient/volumetric level (3D) using different thresholding values for voting at the patient level for the latter. The best results were achieved using the ResNet14 architecture (referred to as AutoML model) via 2D images.

In another prior work, a 3D CNN-based network with BERT was used to classify slices of CT scans [22]. The model used only part of the images from the COV19-CT-DB database. The training and validation set of images were passed through a lung segmentation process first to filter out images of closed lungs and to remove background. After the segmentation process, a resampling method was used to select a set of a fixed number of slices for training and validation. The 3D CNN-based model was followed by a second level MLP classifier to capture all the slices' information from 3D-volumetric images. The final model architecture achieved improved accuracy and macro F1 score on the validation set.

Another study introduced 2D and 3D deep learning models to predict COVID-19 cases [23]. The 2D model, named Deep Wilcoxon signed-rank test (DWCC), adopts non-parametric statistics for deep learning, making the predicted result more stable and explainable, finding a series of slices with the most significant symptoms in a CT scan. On the other hand, the 3D model was based on pixel- and slice-level context mining. The model was termed as CCAT (Convolutional CT scan Aware Transformer), to further explore the intrinsic features in temporal and spatial dimensions.

More work on the same dataset involved deploying a hybrid deep learning framework named CTNet which combines a CNN and a transformer network together for the detection of COVID-19. The method deploys a CNN feature extractor module with Squeeze-and-Excitation (SE) attention to extract features from the CT scans, together with a transformer model to model the discriminative features of the 3D CT scans. The CTNet provides an effective and efficient method to perform COVID-19 diagnosis via 3D CT scans with data resampling strategy. The method's macro F1 score exceeded the baseline on the test partition of the COV19-CT-DB database [24].

Additionally, on the COV19-CT-DB, two experimental methods that customized and combined Deep Neural Network to classify the series of 3D CT-scans chest images were deployed. The proposed methods included experimenting with 2 backbones: DenseNet 121 and ResNet 101. The experiments were separated into 2 tasks, one was for 2 backbones combination of ResNet and DenseNet and the other one was for DenseNet backbones combination. [25] The method's macro F1 score on the test partition of COV19-CT-DB exceeded the baseline model score as can be seen on the leaderboard [26].

The proposed deep learning approaches in the literature summarized above achieved high macro F1 scores on the COV19-CT-DB database. Our work presented here further explores the database and introduces a less hand-engineered and more efficient deep learning based solution for COVID-19 diagnosis. Our model's performance is compared to state-of-the-art on the same dataset.

## 3. METHODOLOGY

### 3.1 The Dataset

COV19-CT-DB is the dataset used for validating the CNN model proposed in this paper as well as other state-of-the-art models compared to it. The CT images in the database were manually annotated by experts and distributed for academic research purposes via the "AI-enabled Medical Image Analysis Workshop and Covid-19 Diagnosis Competition" [27].

The database consists of about 5000 3D chest CT scans acquired from more than 1000 patients. The training set contains 1560 scans in total with 690 of the cases being COVID while the rest (870) belong to the Non-COVID class. The validation set contains, in total, 374, where 165 are COVID cases and 209 are Non-COVID cases. The CT scans in the database contain largely varying slice numbers, ranging from 50 to 700. Please note that the COV19-CT-DB database includes 3 different sets/partitions: a training set, a validation set, and a test set.

The data is unbalanced in terms of the number of 2D slices for both COVID and Non-COVID classes. The images, which are input to the model, were received mainly in loosely compression format; Joint Photographic Experts Group (JPEG) format, grayscale images, with 8-bit depth. The images were all resized to an original size of 512x512 and processed as such.

The numbers of 2D slices used in our work were 335672 in the training set and 75532 in the validation set. Fig. 1 shows distribution of the slices with respect to the classes in the training and validation sets used in this study.

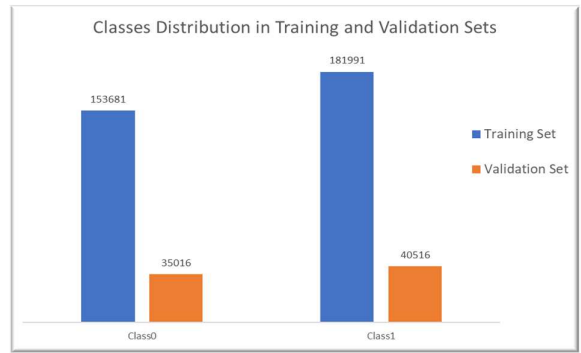

*Fig. 1 Class distribution of the slices in the dataset (Class0 is COVID and Class1 is Non-COVID)*

### 3.2 The Model Architecture

The proposed model's architecture consists of four similar sequential 2D convolutional layers followed by a flatten layer, and two dense layers.

The number of filters in the convolutional layers are 16, 32, 64, and 128, in order, all with a 3x3 filter size. Padding was also applied on the input images in all four convolutional layers, to match input and output image sizes (Padding="same"). The four layers had batch normalization and max polling (2,2), and ReLu (rectified linear unit) activation function with a binary output for the final diagnosis. Fig. 2 shows the proposed CNN model's architecture.

Following the four convolutional layers was a flattening layer, followed by a dense layer with the dimensionality of 256, batch normalization, ReLu activation function, and a dropout of 0.1. The model then ends with a dense layer using a sigmoid activation function.

The model could be replicated in the following sequence:

- Convolutional layer (512, 512,[1] 16) with padding, Batch Norm, Relu activation.
- Max pooling 2D (256, 256, 16),
- Convolutional layer (256, 256, 32) with padding, Batch Norm, Relu activation.
- Max pooling 2D (128, 128, 32)
- Convolutional layer (128, 128, 64) with padding, Batch Norm, Relu activation
- Max pooling 2D (64, 64, 64)
- Convolutional layer (64, 64, 128) with padding, Batch Norm, Relu activation
- Max pooling 2D (32, 32, 128)
- Flatten (131072)
- Dense (256)
- Batch Norm (256)
- ReLu activation (256)
- Dropout (256)

---

[1] 512x512 was the size of the original images in COV19-CT-DB database. This was used to obtain results at slice level. In the next stage, different sizes of images (227x300) were used following slice processing in section 3.3. Thus, the corresponding image sizes of the following layers adjusts accordingly.

- Dense (1)

The output of the final dense layer "Dense (1)" is class1 probability, i.e. the probability of the CNN model predicting class1 corresponding to the Non-COVID class.

The model was compiled using Adaptive momentum estimator (Adam) optimizer with all its default values on Keras [28]. Learning scheduler, learning rate decay and step decay options were not employed. Batch size of 128 is used. The learning rate decay on the original dataset was not employed.

The motivation behind the model architecture is to adopt similar and simple four layers model with standard components: multiple filters, padding, max pooling, and regularization. The four layers are followed by two dense layers and a dropout layer.

### 3.3 Slice Processing

The activation visualization results of classification on the database show room for improvement in terms of accuracy. Following Grad-Cam visualization in Fig. 9, one can theorize that masking the images with the lung area should improve the performance as the model can better learn to discriminate COVID from Non-COVID. To prove the theory and improve the performance, a fixed-sized rectangular Region of Interest (ROI) was applied to localize the anatomy of interest (lung regions) in the slices. The rectangular area was empirically set to contain both left and right lungs over all slices of every scan in the database. Fig. 3 shows this ROI overlaid on an original slice from the training set.

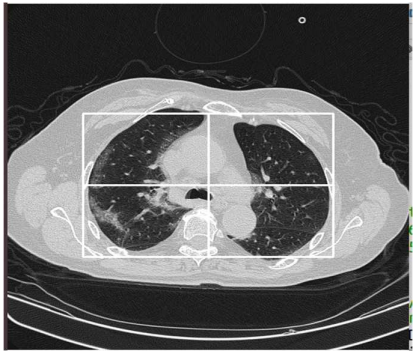

*Fig. 3 Static Rectangular Cropping of images*

After cropping, thresholding was applied to identify and remove uppermost and lowermost slices of the CT scans, corresponding to non-representative slices of the lung volume, aiming to achieve better performance at the patient level diagnosis. Identification of the non-representative slices was realized based on the number of bright pixels in a binarized slice. This procedure is explained below.

First, the cropped images were blurred by using a Gaussian filter to suppress noise and thus enhance large structures in the image. A Gaussian function with a standard deviation of one was convolved with the cropped image's pixel intensity values. The Gaussian function can be expressed in two dimensions as in Equation 1:

$$G(x.y) = \frac{1}{\sqrt{2\pi\sigma^2}} e^{\frac{-x^2-y^2}{2\sigma^2}} \qquad \textit{Equation 1}$$

Where x and y are the distances from the origin in the horizontal and vertical directions, respectively, and $\sigma$ is the standard deviation ($\sigma = 1$).

Second, a histogram based binarization was applied to the resulting blurred images. By looking at the slice's histograms, an estimated threshold for histogram-based image binarization was empirically chosen to be 0.45. This fixed threshold was chosen after applying scale ([0,1]) normalization to the voxel intensities of each scan. Fig. 4 illustrates an exemplary histogram of one of the Gaussian blurred images in the database and the corresponding resulting binarized image.

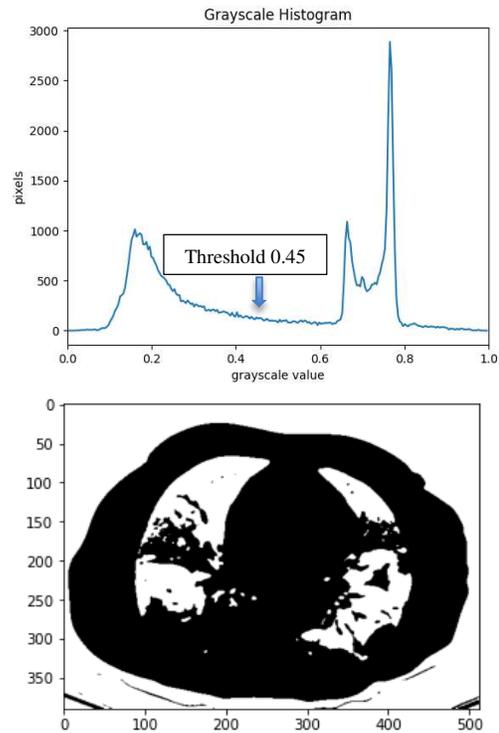

*Fig. 4 Histogram of cropped and blurred image (top) and the resulting binary image (bottom)*

Finally, the binarized image's pixels were used to find a threshold to remove non-representative slices of the CT volume. To choose the threshold, four candidate CT scans were arbitrarily selected from the training set (CT scans 5, 6, 7, and 8) and random slices from them were processed as explained above. To indicate the importance of the slices in the CT volume, labels from one to three were used with three being the most representative/important slice and 1 being the least representative slice; An important slice means a representative slice or a slice that shows a large area of the lung. Similarly, slices of less importance are those that display little to no lung area. Fig. 5 shows the results of the four candidate CT scan volume slices. The chosen filtering threshold for the number of white pixels was 0.066 (corresponding to 4500 out of a total of 68100 pixels in a 227x300 sized slice). Consequently, if the resulting binarized image has more white pixels than the threshold, the slice corresponding to the binarized image will be kept in the CT scan, otherwise it will be removed. The threshold was chosen so as to keep at least one representative slice in every CT scan volume to be part of the final diagnosis.

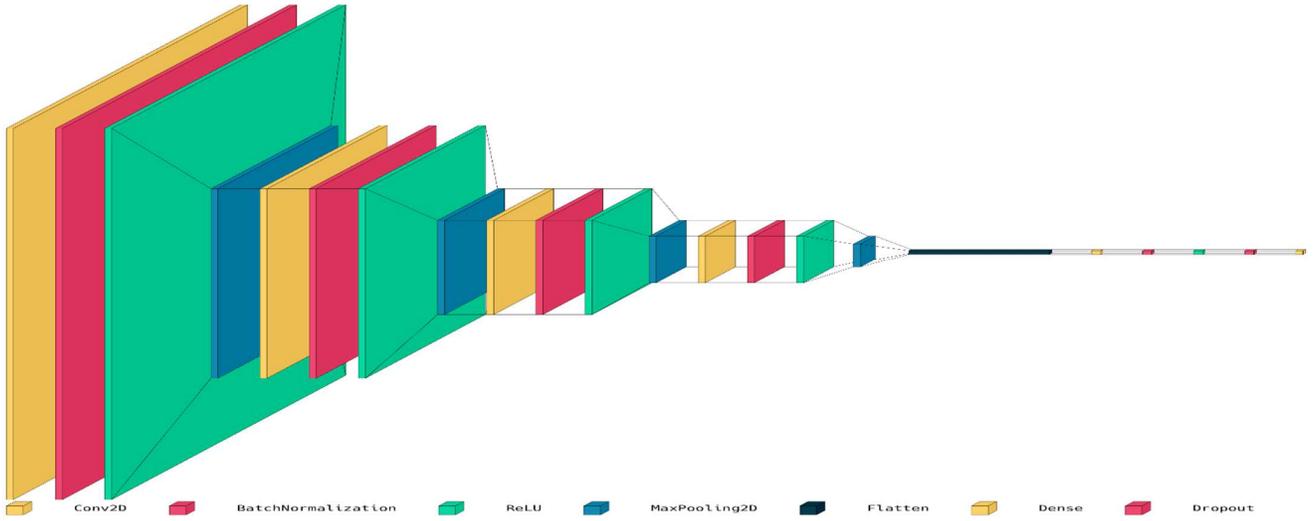

*Fig. 2 Visualization of the layers of the proposed CNN model*

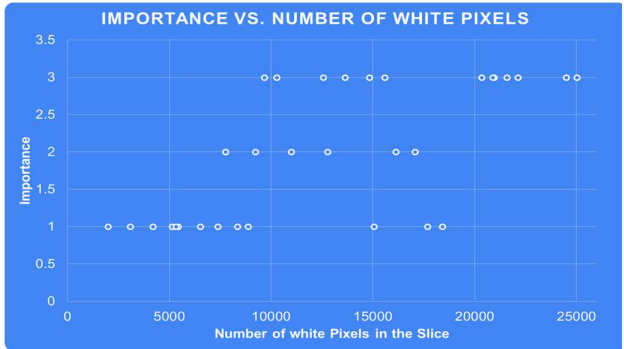

*Fig. 5 Number of white pixels in the candidate CT volume's slices against the importance of the slice in the CT volume*

The slice processing methodology reduces the number of slices in the dataset by including only the representative slices. Accordingly, the number of training and validation slices were reduced to 280462 (corresponding to 16% reduction) and 63559 (15.3% reduction), respectively. Please note that the original number of the slices are as shown in Fig. 1 above.

### 3.4 Hyperparameters Tuning

After slice processing, hyperparameters were tuned using the same CNN model architecture.

To stabilize the increment in validation accuracy during training, a learning rate scheduler was used to prevent fluctuation. The learning rate scheduler included an exponential decay function, applied to the SGD (Stochastic Gradient Decent) optimizer step. This function can be useful for changing the learning rate value across different invocations of optimizer functions. The decay of the learning is computed as in Equation 2:

$$\text{Learning Rate} = \text{Initial LR} \times \text{Decay Rate}^{\frac{Optimizer\ step}{Decay\ step}}$$

*Equation 2*

The initial learning rate (initial LR) was set to 0.1 and a 0.96 decay rate was used. The value of steps divided by decay steps is an integer division, i.e. the decayed learning rate follows a staircase function.

The optimizer's steps were defined using floor divisions as in Equation 3:

$$train_{step} = \lfloor number\ of\ train\ samples/batch\ size \rfloor = \lfloor 280462/128 \rfloor$$

$$validation_{step} = \lfloor number\ of\ (test)\ validation\ samples\ /batch\ size \rfloor = \lfloor 63559/128 \rfloor$$

*Equation 3*

Decay steps were set every 100000 steps [29].

On the other hand, class weights were used to modify any imbalance in the input image classes. The ratio of the class weight in our case study came to {1.197:1} COVID to Non-COVID ratio in the training set. The class weight was calculated using the formula in Equation 4:

$$class\ weight = \frac{number\ of\ images\ for\ a\ class\ in\ the\ training\ set}{number\ of\ all\ images\ in\ the\ training\ set}$$

*Equation 4*

Finally, image augmentation (mainly horizontal and vertical flipping) was applied on the processed. These image flipping techniques aimed at improving the accuracy via smoothing the effects of the content variations present in the slice [30].

### 3.5 Patient Level Decision

At the patient level, different class probability thresholds were tried and compared using class prediction probability to achieve the highest diagnosis accuracy. The class probability thresholds were based on the probability of prediction of class 1 (Non-COVID); If the output probability for class 1 is greater than the chosen threshold, then the slice would be predicted as Non-COVID. Otherwise, the slice would be predicted as COVID. In that, if number of COVID slices is equal to the number of Non-COVID slices in any one of the CT volume, then the decision is that the patient is a Non-COVID. This slice level decision can be expressed as follows:

    if Class1 probability > class probability threshold:

     Predict slice as Non-COVID

    else:

     Predict slice as COVID

After slice level predictions are obtained, a patient is diagnosed based on the presence/absence of COVID slices in his/her CT: if patient CT data contains more Non-COVID predicted slices than COVID predicted slices, the patient is diagnosed as Non-COVID else the patient is diagnosed as COVID (majority voting method).

The clinical relevance of the patient level diagnosis approach we presented above can be explained as follows. Assuming that a patient has lung damage due to COVID seen in 30% of its slices. So, our network classifies around 30% of the slices as COVID and the rest as Non-COVID, and thus the final result will be Non-COVID (in line with majority voting).

While even a minor anomaly seen in a single slice may be attributed to a disease, we speculate that in the Covid case a reasonable amount of involvement is necessary for the diagnostic decision to be taken, and our deep learning model is highly sensitive to even the smallest anomalies observed in the slices.

Please note that we also tried the "all-or-nothing approach" where COVID diagnosis decision is taken even a single slice is predicted as COVID, but that approach yielded less accurate results - as elaborated in the results section - supporting our above observations.".

### 3.6 Performance Evaluation

The proposed model was evaluated via the COV19-CT-DB database. The macro F1 score was calculated after averaging precision and recall matrices as in Equation 5:

$$macro\ F1 = \frac{2 \times average\ precision \times average\ recall}{average\ precision + average\ recall}$$

*Equation 5*

Average precision and average recall are measured on the validation set.

Furthermore, in an attempt to report the confidence intervals of the results obtained, the Binomial proportion confidence intervals for macro F1 score are used. The confidence intervals were calculated from the following formulation [31]:

The radius of the interval is defined as in Equation 6:

$$radius\ of\ interval = z \times \sqrt{\frac{macro\ F1 \times (1 - macro\ F1)}{n}}$$

*Equation 6*

Where $n$ is the number of samples used.

In the above formulation, z is the number of standard deviations from the Gaussian distribution, which is taken as z=1.96 for a significance level of 95%.

Performance evaluation of our method is conducted at slice-level and at patient-level, where the former corresponds to considering 2D slices individually in any quantitative or qualitative analysis. Whereas, in patient level results CT volumes are considered as a whole, and thus the prediction is emphasizing 3D-CT prediction value or patient level rather than each 2D slice's predicted value.

## 4. RESULTS

### 4.1 Slice Level Results

The effect of increasing batch size on the diagnostic performance has been exploited. In that, using a batch size of 64 sufficiently improved the performance as compared to smaller batch sizes. The macro F1 score increased from 0.903 with a batch size of 32 to 0.927 with a batch size of 64. Finally, using 128 as a batch size increased the resulting macro F1 to the number reported in the results section. Accordingly, a 128 batch size was used in the rest of the experiments.

As the batch size increased the computation time also increased. Training the CNN model using a batch size of 128 took about two and half days over a workstation using GNU/Linux operating system on 62GiB System memory with Intel(R) Xeon(R) W-2223 CPU @ 3.60GHz processor.

Our model achieved average recall and precision rates of 0.95 and 0.93 on the validation set, respectively. The training was conducted over 70 epochs, which was empirically selected by monitoring the performance metrics on different partitions over sufficient number of epochs to avoid overtraining and undertraining. With these rates, the macro F1 score reached 0.94 for this binary classification, which was obtained using the CNN model on the original images in the database, i.e. without any slice processing, or hyperparameters tuning. Adam optimizer was used in these results. Using other optimizer options, such as Stochastic SGD, did not yield higher macro F1 score. Moreover, adding a learning scheduler or employing step decay, was also tried and found to add only fractional improvement in the results. Fig. 6 shows the evolution of recall and precision rates on the validation set over the epochs. The images were used in their original sizes.

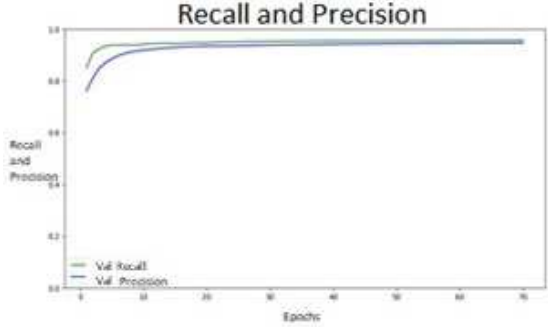

*Fig. 6 Evolution of recall and precision rates on the validation set for the CNN model*

The interval of the reported macro F1 score with 95% significance level is calculated as in Equation 7:

$$interval = 1.96 \times \sqrt{\frac{0.94\,(1-0.94)}{75532}} \approx 0.0017$$

*Equation 7*

The results show a narrow deviation from our reported macro F1 score.

The proposed model was compared to other state-of-the-art models on the COV19-CT-DB database in terms of the macro F1 score and the confidence intervals. Our proposed CNN model, although has a simple architecture, achieved improved performance in terms of macro F1 score with similar or better confidence intervals compared to the other models. Table I shows a comparison of the proposed model and the state-of-the-art models mentioned in the related work section. The comparison is realized both at patient level and at slice level for the methods presented. As observed, the results encourage using the CNN model at the patient level with 128 batch size. The accuracy of the model is considered next.

Table I. Comparison of the proposed model with the state-of-the-art and the baseline model on COV19-CT-DB validation set

| The Model | Macro F1 | Confidence Intervals |
|---|---|---|
| ResNet50-GRU (Baseline model) [17] | 0.70 | ± 0.0032 |
| AutoML model (*best-reported result*) [21] | 0.88 | ± 0.0023 |
| 3D-CNN-Network with BERT [23] | 0.92 | ± 0.0018 |
| CCAT and DWCC [24] | 0.93 | ± 0.0017 |
| Our proposed Methodology | 0.94 | ± 0.0017 |

Our proposed model reached a validation accuracy of about 80% on the original dataset. Fig. 7 shows the training and validation accuracies. The model allows sharp learning during the initial epochs and a steadier trend throughout the rest of the epochs. Furthermore, hyperparameters tuning was not used to train the model and that could explain the fluctuation and spikes of the results on the validation set.

In terms of model generalizability, we anticipated that the CNN model architecture and the large dataset we used would prevent overfitting at the expense of longer training (70 training epochs). The dataset was large as discussed in Section 3.1. Moreover, the CNN model included relatively few numbers of hidden layers (only 4). Indeed, the training accuracy does not reach 100% but keeps nearing it at later epochs. Similarly, the validation accuracy frequents around a percentage less that the training accuracy. Practically, however, in terms of the macro F1 score (recall and precision), the CNN model performed distinctively well on the training and the validation sets as illustrated in Fig. 6 before, and the validation accuracy averaged about 80%. Furthermore, the same model architecture with tunned parameters performed similarly well on unseen images (the test partition).

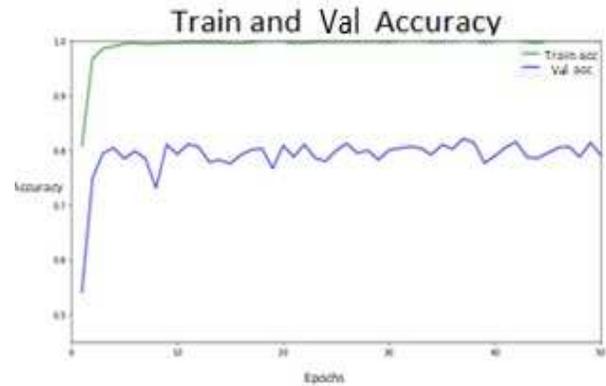

*Fig. 7 Evolution of the training and validation accuracies for the proposed model.*

On the other hand, to understand how our proposed model performs the classification, Guided Grad-cam class activation visualization was used at the last convolutional layer of the model - the layer followed by a (256) flatten layer [32]. Fig. 8 shows the Grad-cam visualization for a slice in the validation set. The color map used in the figure is the Viridis colormap [33]. The slice belongs to a COVID case and was correctly classified by the model. The outputs for the correct and incorrect classifications are adapted to the input image. They clearly show that the model pays attention to:

- the lung area, and

- the posterior and anterior walls (with the anterior walls getting very strong attention values).

We can observe a similar attention distribution on the COVID cases incorrectly classified and Non-COVID cases (correct and incorrectly classified slices) as well.

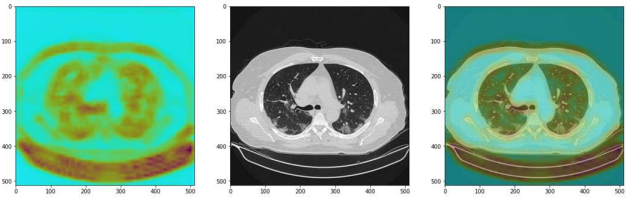

*Fig. 8 Guided Grad-Cam visualization (left) of a correctly predicted COVID slice (middle). The image on the right pane displays an overlaid version of the two.*

As for the slice level decision, the proposed model can sometimes incorrectly predict the uppermost and the lowermost slices as Non-COVID (specifically, 20 out of 24 extreme slices in the validation partition are misclassified). These extreme slices correspond to the anatomical regions where COVID involvement is not seen, and therefore can be considered the least representative slices for the diagnosis of the disease. Fig. 9 shows exemplary slices that are correctly classified by our proposed model, while Fig. 10 depicts exemplary slices that are incorrectly classified where the extreme slices can be observed.

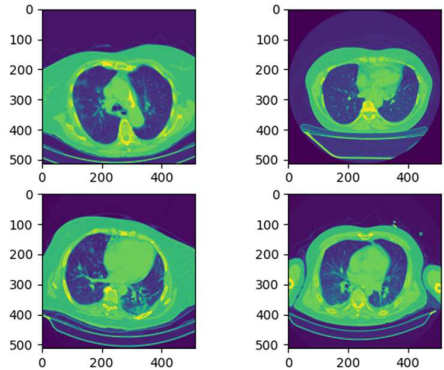

*Fig. 9 Examples of correctly classified slices from COVID (right) and Non-COVID (left) cases*

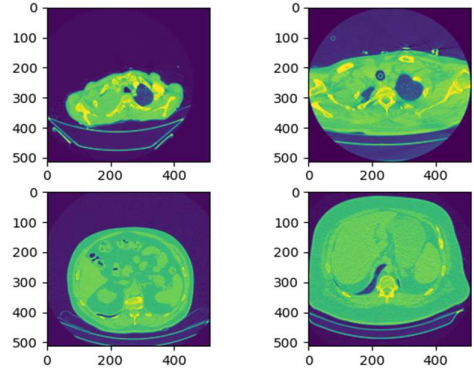

*Fig. 10 Examples of incorrectly classified slices from COVID (right) and Non-COVID (left) cases*

To increase the validation accuracy, the slices were processed as described in Section 3.3 and the parameters were tuned as described in Section 4.3. With that, the CNN model reached a validation accuracy of 84%, improved from 80% validation accuracy achieved without slice processing and hyperparameters tuning. The final model including slice processing and hyperparameters tuning was used for taking patient diagnosis.

### 4.2 Patient Level Results

In order to obtain patient level diagnosis from slice level decisions different class probability thresholds (for slices prediction) varying in the range of [0,1] were tried as explained in Section 3.5, and the corresponding macro F1 scores were compared. Majority voting was used at patient level (for CT prediction). As observed in Fig. 11, the model achieves the highest macro F1 score with a class probability threshold of 0.40, followed by class probability threshold of 0.15. The validation accuracies when using the mentioned thresholds are 88.5% and 87.7%, respectively. With that, the results demonstrate that a class probability threshold of 0.40 achieves the best performance when used with majority voting at the patient level in terms of macro F1 score compared to the other class probability threshold values. The patient level macro F1 score achieved using the proposed method reaches 0.882 on the validation set. The resulting macro F1 score of the proposed model comfortably exceeds that of the baseline model on the validation set. The baseline score on the COV19-CT-DB validation set is 0.70, as reported in [17].

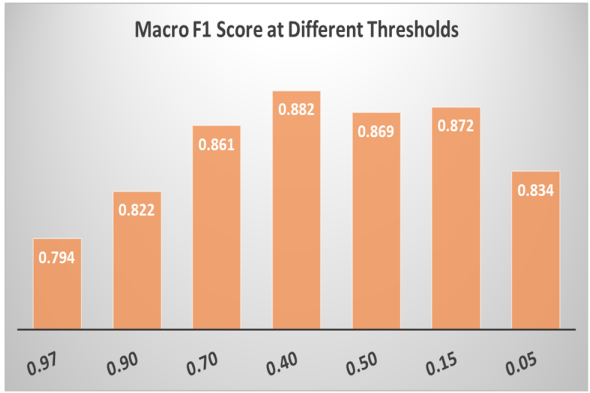

*Fig. 11 Different class probability thresholds (horizontal axis) for making predictions at patient level and the corresponding macro F1 scores.*

In general, the model misclassifies 13 Non-COVID cases out of 209, and 30 COVID ones out of 165. Class-specific macro F1 scores of the proposed method are 0.86 for the COVID class and 0.90 for the Non-COVID. Table II shows the confusion matrix of the proposed method at the patient level for the best threshold value.

Table II. Confusion matrix on patient level decision using a voting threshold of 0.40. Columns refer to actual cases while rows display predictions of the proposed model.

|  | Actual | |
|---|---|---|
|  | COVID | Non-COVID |
| COVID | 135 | 13 |
| Non-COVID | 30 | 196 |

Despite the fact that misclassification of 13 Non-COVID cases out of 209 is less problematic than misclassification of 30 COVID cases out of 165, the model aimed mainly at increasing the quantitative results. Further, the work here considers automation of the solutions are equal as diagnostic accuracy.

On the other hand, a method of deciding a one CT scan to be COVID based on finding at least one COVID slice, named 'any covid gives covid' was tried. The aim was decreasing misclassification of COVID cases. However, even though the COVID cases misclassification was decreased from 30 to 16 misclassification cases from the above-mentioned method, the macro F1 score gave lower performance. The validation accuracy was 77.2% and the macro F1 score was 0.720 less than 0.86 of the above-mentioned method. Table III shows the confusion matrix resulting from "any covid gives covid" method.

Table III. Confusion matrix on patient level decision using any covid gives covid method with a voting class probability threshold of 0.40. Columns refer to actual cases while rows display predictions of the proposed model.

|  | Actual | |
|---|---|---|
|  | COVID | Non-COVID |
| COVID | 149 | 88 |
| Non-COVID | 16 | 196 |

Further, to validate the results the method was tested on the test partition of the COV19-CT-DB database (unseen images). On unseen dataset of images, the method's performance exceeded those of the baseline and other works. Within the context of the MIA-COVID19 competition, the teams were provided with a test partition of images. The model achieved 0.82 macro F1 score, with 0.96 F1 score for Non-COVID and 0.68 F1 score for COVID. This score is above the baseline, which is 0.67 macro F1 score.

Our proposed method did not only exceed the baseline macro F1 score (0.67), but also outperformed other alternatives entered the competition and reported accuracies on COV19-CT-DB's test partition [26]. Table IV compares our model to other alternatives on the test partition (unseen images) of COV19-CT-DB. Our team is named "IDU-CVLab" and the code was developed using python.[2]

Table IV. Comparison of the proposed method with the base line and other alternatives on the COV19-CT-DB test partition (unseen images)

| The Model | Macro F1 |
|---|---|
| ResNet50-GRU (Baseline model) [17] | 0.67 |
| A hybrid deep learning framework (CTNet) [24] | 0.78 |
| Custom Deep Neural Network [25] | 0.78 |
| Our proposed methodology (IDU-CVLab) | 0.82 |
| CCAT and DWCC [24] | 0.88 |

## 5. CONCLUSION AND DISCUSSION

This paper proposes a solution for COVID-19 diagnosis using deep learning and image processing techniques. The adopted CNN model architecture was trained on the training set of COV19-CT-DB and validated by the slice level prediction performance at the COV19-CT-DB validation set. The model's macro F1 score was compared to those of other alternatives on the same validation set. The model achieved a state-of-the-art macro F1 score with similar or better confidence intervals for classification on the validation set (COV19-CT-DB validation set) at slice level. Further, the validation accuracy of the model was improved by performing a slice processing along with model parameters at patient level. The new method was tested with different class probability thresholds to select the best probability threshold for each 3D-CT scan predictions via the validation set.

---
[2] https://github.com/IDU-CVLab/COV19D

Next and to make predictions at patient level, the final method was implemented using majority voting on each CT scan. The final method achieved a macro F1 score exceeding the base line score and other alternatives on the test partition of the COV19-CT-DB database.

More complex modelling techniques do not reach as high macro F1 scores as the CNN model trained in this paper at slice level. With that, the paper generally encourages researchers, programmers, and otherwise to consider a simpler and from scratch deep learning model with different modifications, rather than using pretrained models or similar method.

On the other hand, using the rectangular region selection for slice processing improved the method's performance or the accuracy. This goes to prove that limiting the region of interest with the lung volumes instead of processing the whole CT scan will be a promising approach. Therefore, segmenting lung parenchyma prior to classification could further improve diagnostic performance of the proposed method.

## ACKNOWLEDGEMENT

The authors acknowledge the work of all the medical staff and others who manually annotated the images in the COV19-CT-DB database and shared them in a relatively big dataset.

Funding: This research did not receive any specific grant from funding agencies in the public, commercial, or not-for-profit sectors.